\documentclass[aps,prd,twocolumn,groupedaddress,showpacs,amsmath,nofootinbib]{revtex4}
\usepackage{graphicx}

\begin{document}

\title{Dark energy and dark matter as curvature effects}

\author{S. Capozziello$^{\diamond}$}
\author{V.F. Cardone$^{\natural}$}
\author{A. Troisi$^{\diamond}$} \affiliation{$^\diamond$
Dipartimento di Scienze Fisiche, Universit\'{a} di Napoli
``Federico II", INFN, Sez. di Napoli, Compl. Univ. di Monte
S.Angelo, Edificio G, Via Cinthia, I\,-\,80126\,-\, Napoli, Italy
 \\
$^\natural$  Dipartimento di Fisica ``E.R. Caianiello",
Universit\`{a} di Salerno,
 Via S. Allende, I\,-\,84081\,-\,Baronissi (SA), Italy}

\begin{abstract}

Astrophysical observations are pointing out huge amounts of "dark
matter" and "dark energy" needed to explain the observed large
scale structures and cosmic accelerating expansion. Up to now, no
experimental evidence has been found, at fundamental level, to
explain such mysterious components. The problem could be
completely reversed considering dark matter and dark energy as
"shortcomings" of General Relativity and claiming for the
"correct" theory of gravity as that derived by matching the
largest number of observational data. As a result, accelerating
behavior of cosmic fluid and rotation curves of spiral galaxies
are reproduced by means of "curvature effects".

\end{abstract}


\maketitle

The impressive amount of unprecedented quality data of last decade
has made it possible to shed new light on the knowledge of the
Universe.  Type Ia Supernovae (SNeIa),  anisotropies in the cosmic
microwave background radiation (CMBR), and matter power spectrum
inferred from large galaxy surveys  represent the strongest
evidences for a radical revision of the Cosmological Standard
Model. In particular, the {\it concordance $\Lambda$CDM model}
predicts that baryons contribute only for $\sim 4\%$ of the total
matter\,-\,energy budget, while the exotic {\it cold dark matter}
(CDM) represents the bulk of the matter content ($\sim 25\%$) and
the cosmological constant $\Lambda$ plays  the role of the so
called dark energy ($\sim 70\%$). Although being the best fit to a
wide range of data, the $\Lambda$CDM model is severely affected by
strong theoretical shortcomings  that have motivated the search
for alternative approaches. Dark energy models mainly rely on the
implicit assumption that Einstein's General Relativity is indeed
the correct theory of gravity. Nevertheless,  it is  conceivable
that both cosmic speed up and dark matter represent signals of a
breakdown in our understanding of the laws of gravitation so that
one should consider the possibility that the Hilbert\,-\,Einstein
Lagrangian, linear in the Ricci scalar $R$, should be generalized.
Following this approach, the choice of the effective gravity
Lagrangian can be derived by means of the data and the "economic"
requirement that no exotic ingredients have to be added. This is
the underlying philosophy of what is referred to as $f(R)$ gravity
\cite{capozzcurv,cdtt,flanagan,francaviglia,odintsovfr}.  From a
theoretical standpoint, several issues from fundamental physics
(quantum field theory on curved spacetimes, M-theory etc.) suggest
that higher order terms must necessarily enter the gravity
Lagrangian. On the other side,  Solar System experiments show the
validity of Einstein's theory at these scales. In other words, the
PPN limit of $f(R)$-models  must not violate the experimental
constraints on Eddington parameters. A positive answer to this
request has been recently achieved for several $f(R)$ theories
\cite{ppnantro}, nevertheless it has to be remarked that this
debate is far to be definitively concluded. Although higher order
gravity theories have received much attention in cosmology, since
they are naturally able to give rise to accelerating expansions
(both in the late and the early universe
\cite{starobinsky,kerner}), it is possible to demonstrate that
$f(R)$ theories can also play a major role at astrophysical
scales. In fact, modifying the gravity Lagrangian can affect the
gravitational potential in the low energy limit \cite{stelle},
provided that the modified potential reduces to the Newtonian one
on the Solar System scale.  In fact, a corrected gravitational
potential could offer the possibility to fit galaxy rotation
curves without the need of dark matter \cite{noipla,mond}. In
addition, one could work out a formal analogy between the
corrections to the Newtonian potential and the usually adopted
dark matter models.    The choice of an analytic function in term
of Ricci scalar is physically corroborated by the Ostrogradski
theorem \cite{woodard}, which states that this kind of Lagrangian
is the only viable one which can be considered among the several
that can be constructed by means of curvature tensor and possibly
its covariant derivatives. The field equations of this approach,
recast in the Einstein form, read
 \cite{capozzcurv}\,:
\begin{equation}\label{5}
G_{\alpha \beta} = R_{\alpha\beta}-\frac{1}{2}g_{\alpha\beta}R =
T^{curv}_{\alpha\beta}+T^{M}_{\alpha\beta}/f^\prime(R)
\end{equation}
where the prime denotes derivative with respect to $R$,
$T^{M}_{\alpha \beta}$ the standard matter stress\,-\,energy
tensor and
\begin{displaymath}
T^{curv}_{\alpha\beta}\,=\,\frac{1}{f'(R)}\Big\{\frac{1}{2}g_{\alpha\beta}\left[f(R)-Rf'(R)\right]
+
\end{displaymath}
\begin{equation} \label{6}
f'(R)^{;\mu\nu}(g_{\alpha\mu}g_{\beta\nu}-g_{\alpha\beta}g_{\mu\nu})
\Big\}\,,
\end{equation}
defines the {\it curvature stress\,-\,energy tensor}. The presence
of the terms $f^\prime(R)_{;\mu\nu}$ renders the equations of
fourth order, while, for $f(R) = R$,  Eqs.(\ref{5}) reduce to the
standard second\,-\,order Einstein field equations. As it is clear
from Eq.(\ref{5}), the curvature stress\,-\,energy tensor formally
plays the role of a further source term in the field equations
which effect is the same as that of an effective fluid of purely
geometrical origin. Depending on the scales, it is such a
curvature fluid which can play the role of dark matter and dark
energy. From the cosmological viewpoint, in the standard framework
of a spatially flat homogenous and isotropic Universe, the
cosmological dynamics is determined by its energy budget through
the Friedmann equations. In particular, the cosmic acceleration is
achieved when the r.h.s. of the acceleration equation remains
positive. In physical units, we have $  \ddot{a}/{a} = - (1/6)  (
\rho_{tot} + 3 p_{tot}  ) \,,$ where $a$ is the scale factor, $H =
\dot{a}/a$ the Hubble parameter, the dot denotes derivative with
respect to cosmic time, and the subscript $tot$ denotes the sum of
the curvature fluid and the matter contribution to the energy
density and pressure. From the above relation, the acceleration
condition, for a dust dominated model, leads to:
$\rho_{curv} + \rho_M + 3p_{curv} < 0 \rightarrow w_{curv} < -
\frac{\rho_{tot}}{3 \rho_{curv}}$
so that a key role is played by the effective quantities\,:

\begin{equation}
\rho_{curv} = \frac{1}{f'(R)} \left \{ \frac{1}{2} \left [ f(R)  -
R f'(R) \right ] - 3 H \dot{R} f''(R) \right \} \ , \label{eq:
rhocurv}
\end{equation}
\begin{equation}
w_{curv} = -1 + \frac{\ddot{R} f''(R) + \dot{R} \left [ \dot{R}
f'''(R) - H f''(R) \right ]} {\left [ f(R) - R f'(R) \right ]/2 -
3 H \dot{R} f''(R)} \ . \label{eq: wcurv}
\end{equation}
As a direct simplest choice, one may assume a power\,-\,law form
$f(R) = f_0 R^n$, with $n$ a real number, which represents a
straightforward generalization of the Einstein General Relativity
in the limit $n=1$. One can find power\,-\,law solutions for
$a(t)$ providing a satisfactory fit to the SNeIa data and a good
agreement with the estimated age of the Universe in the range
$1.366 < n < 1.376$ \cite{curv-ijmpd}. It is worth noting, that
even an inverse approach for the choice of $f(R)$ is in order.
Cosmological equations derived from (\ref{5}) can be reduced to a
linear third order differential equation for the function
$f(R(z))$, where $z$ is the redshift. The Hubble parameter $H(z)$
inferred from the data and the relation between $z$ and $R$ can be
used to finally work out $f(R)$ \cite{mimicking}. In addition, one
may consider the expression for $H(z)$ in a given dark energy
model as the input for the above reconstruction of $f(R)$ and thus
work out a $f(R)$ theory giving rise to the same dynamics as the
input model. This suggests the intriguing possibility of
considering observationally viable dark energy models (such as
$\Lambda$CDM and quintessence) only as effective parameterizations
of the curvature fluid \cite{mimicking}.

The successful results obtained at cosmological scales motivates
the investigation of $f(R)$ theories even at astrophysical scales.
 In the low energy limit,
higher order gravity  implies a modified gravitational potential.
Now, by considering the case of a pointlike mass $m$ and solving
the vacuum field equations for a Schwarzschild\,-\,like metric
\cite{noipla}, one gets from a theory $f(R)=f_0 R^n$ the modified
gravitational potential:
\begin{equation}
\Phi(r) = - \frac{G m}{r} \left [ 1 + \left ( \frac{r}{r_c} \right
)^{\beta} \right ] \label{eq: pointphi}
\end{equation}
where
\begin{equation}
\beta = \frac{12n^2 - 7n - 1 - \sqrt{36n^4 + 12n^3 - 83n^2 + 50n +
1}}{6n^2 + 4n - 2} \label{eq: bnfinal}
\end{equation}
which corrects the ordinary Newtonian potential by a power\,-\,law
term. In particular, this correction sets in on scales larger than
$r_c$ which value depends essentially on the mass of the system.
The corrected potential (\ref{eq: pointphi}) reduces to the
standard $\Phi \propto 1/r$ for $n=1$ as it can be seen from the
relation (\ref{eq: bnfinal}). The generalization of Eq.(\ref{eq:
pointphi}) to extended systems is straightforward. We simply
divide the system in infinitesimal mass elements and sum up the
potentials generated by each single element. In the continuum
limit, we replace the sum with an integral over the mass density
of the system taking care of eventual symmetries of the mass
distribution. Once the gravitational potential has been computed,
one may evaluate the rotation curve $v_c^2(r)$ and compare it with
the data. For
 the pointlike case we have\,:
\begin{equation}
v_c^2(r) = \frac{G m}{r} \left [ 1 + (1 - \beta) \left (
\frac{r}{r_c} \right )^{\beta} \right ] \ . \label{eq: vcpoint}
\end{equation}
Compared with the Newtonian result $v_c^2 = G m/r$, the corrected
rotation curve is modified by the addition of the second term in
the r.h.s. of Eq.(\ref{eq: vcpoint}). For $0 <\, \beta \,< 1$, the
corrected rotation curve is higher than the Newtonian one. Since
measurements of spiral galaxies rotation curves signals a circular
velocity higher than what is predicted on the basis of the
observed luminous mass and the Newtonian potential, the above
result suggests the possibility that our modified gravitational
potential may fill the gap between theory and observations without
the need of additional dark matter. It is worth noting that the
corrected rotation curve is asymptotically vanishing as in the
Newtonian case, while it is usually claimed that observed rotation
curves are flat. Actually, observations do not probe $v_c$ up to
infinity, but only show that the rotation curve is flat within the
measurement uncertainties up to the last measured point. This fact
by no way excludes the possibility that $v_c$ goes to zero at
infinity.
\begin{table}
\begin{tabular}{|c|c|c|c|c|c|c|}
\hline Id & $\beta$ & $\log{r_c}$ & $f_g$ & $\Upsilon_{\star}$ &
$\chi^2/dof$ & $\sigma_{rms}$ \\
\hline UGC 1230 & 0.608 & -0.24 & 0.26 & 7.78 & 3.24/8 & 0.54 \\
UGC 1281 & 0.485 & -2.46 & 0.57 & 0.88 & 3.98/21 & 0.41 \\ UGC
3137 & 0.572 & -1.97 & 0.77 & 5.54 & 49.4/26 & 1.31 \\ UGC 3371 &
0.588 & -1.74 & 0.49 & 2.44 & 0.97/15 & 0.23 \\ UGC 4173 & 0.532 &
-0.17 & 0.49 & 5.01 & 0.07/10 & 0.07 \\ UGC 4325 & 0.588 & -3.04 &
0.75 & 0.37 & 0.20/13 & 0.11 \\ NGC 2366 & 0.532 & 0.99 & 0.32 &
6.67 & 30.6/25 & 1.04 \\ IC 2233 & 0.807 & -1.68 & 0.62 & 1.38 &
16.29/22 & 0.81\\ NGC 3274 & 0.519 & -2.65 & 0.72 & 1.12 &
19.62/20 & 0.92 \\ NGC 4395 & 0.578 & 0.35 & 0.17 & 6.17 &
34.81/52 & 0.80 \\ NGC 4455 & 0.775 & -2.04 & 0.88 & 0.29 &
3.71/17 & 0.43 \\ NGC 5023 & 0.714 & -2.34 & 0.61 & 0.72 &
13.06/30 & 0.63 \\ DDO 185 & 0.674 & -2.37 & 0.90 & 0.21 & 6.04/5 & 0.87 \\
DDO 189 & 0.526 & -1.87 & 0.69 & 3.14 & 0.47/8 & 0.21 \\ UGC 10310
& 0.608 & -1.61 & 0.65 & 1.04 & 3.93/13 & 0.50 \\ \hline
\end{tabular}
\caption{Best fit values of the model parameters from maximizing
the joint likelihood function ${\cal{L}}(\beta, \log{r_c}, f_g)$.
We also report the value of $\Upsilon_{\star}$, the $\chi^2/dof$
for the best fit parameters (with $dof = N - 3$ and $N$ the number
of datapoints) and the root mean square $\sigma_{rms}$ of the fit
residuals.}\label{tab}
\end{table}
In order to observationally check the above result, we have
considered a sample of LSB galaxies with well measured HI +
H$\alpha$ rotation curves extending far beyond the visible edge of
the system. LSB galaxies are known to be ideal candidates to test
dark matter models since, because of their high gas content, the
rotation curves can be well measured and corrected for possible
systematic errors by comparing 21\,-\,cm HI line emission with
optical H$\alpha$ and ${\rm [NII]}$ data. Moreover, they are
supposed to be dark matter dominated so that fitting their
rotation curves without this elusive component is a strong
evidence in favor of any successful alternative theory of gravity.
Our sample contains 15 LSB galaxies with data on both the rotation
curve, the surface mass density of the gas component and
$R$\,-\,band disk photometry extracted from a larger sample
selected by de Blok \& Bosma \cite{dbb02}. We assume the stars are
distributed in an infinitely thin and circularly symmetric disk
with surface density $\Sigma(R) = \Upsilon_\star I_0
exp{(-R/R_d)}$ where the central surface luminosity $I_0$ and the
disk scalelength $R_d$ are obtained from fitting to the stellar
photometry. The gas surface density has been obtained by
interpolating the data over the range probed by HI measurements
and extrapolated outside this range.

When fitting to the theoretical rotation curve, there are three
quantities to be determined, namely the stellar
mass\,-\,to\,-\,light (M/L) ratio, $\Upsilon_{\star}$ and the
theory parameters $(\beta, r_c)$. It is worth stressing that,
while fit results for different galaxies should give the same
$\beta$, $r_c$  must be set on a galaxy\,-\,by\,-\,galaxy basis.
However, it is expected that galaxies having similar properties in
terms of mass distribution have similar values of $r_c$ so that
the scatter in $r_c$ must reflect somewhat that on the terminal
circular velocities. In order to match the model with the data, we
perform a likelihood analysis determining for each galaxy using as
fitting parameters $\beta$, $\log{r_c}$ (with $r_c$ in kpc) and
the gas mass fraction\footnote{This is related to the $M/L$ ratio
as $\Upsilon_{\star} = [(1 - f_g) M_{g}]/(f_g L_d)$ with $M_g =
1.4 M_{HI}$ the gas (HI + He) mass, $M_d = \Upsilon_{\star} L_d$
and $L_d = 2 \pi I_0 R_d^2$ the disk total mass and luminosity.}
$f_g$.  Considering the results  summarized in Table \ref{tab},
the experimental data are successfully fitted by the model. In
particular, for the best fit range of $\beta$ $(\beta=0.58\pm
0.15)$, one can conclude that $R^n$ gravity with $1.34 < n <2.41$
(which well overlaps the above mentioned range of $n$ interesting
in cosmology) can be a good candidate to solve the missing matter
problem in LSB galaxies without any dark matter.
\begin{figure}
\centering\resizebox{4cm}{!}{\includegraphics{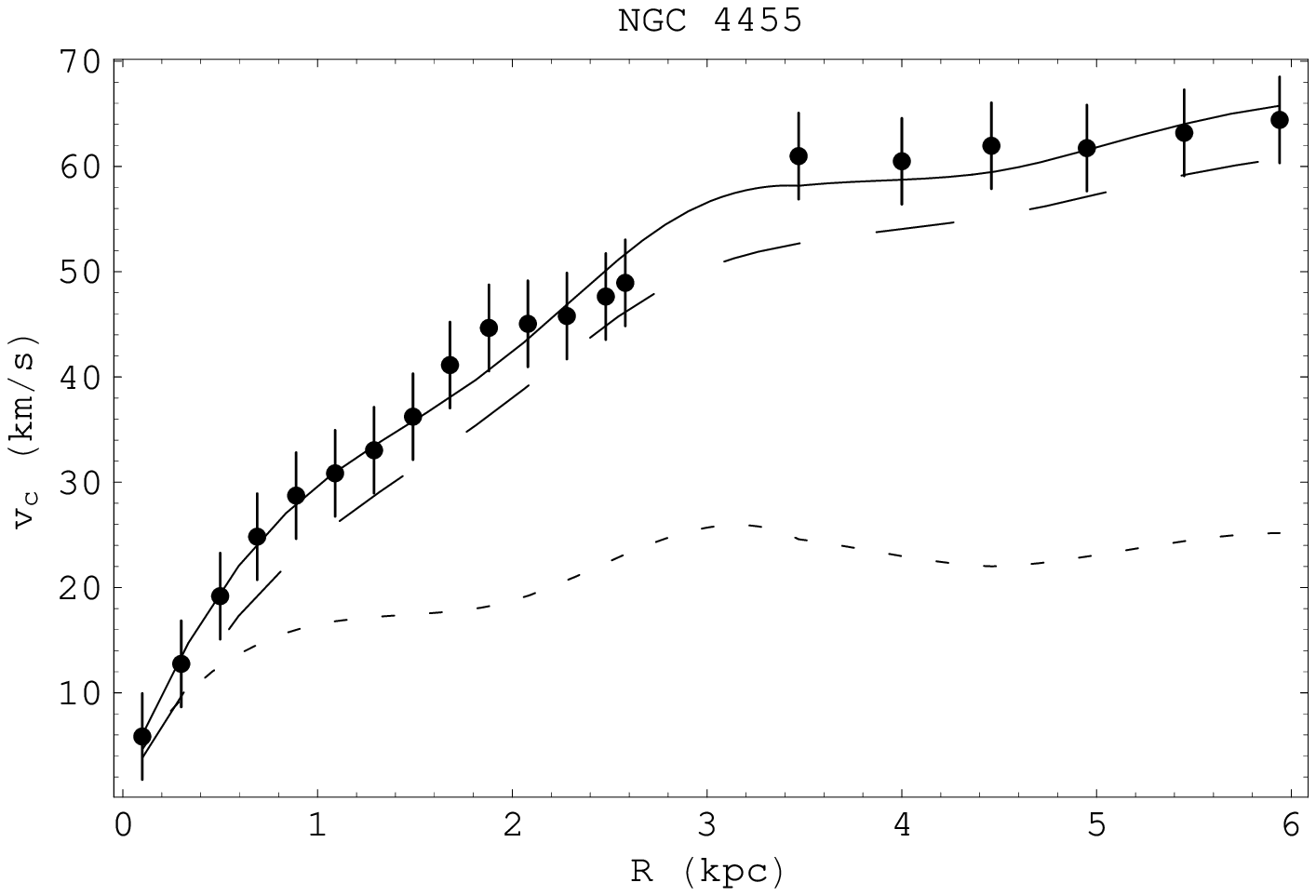}}
\centering\resizebox{4cm}{!}{\includegraphics{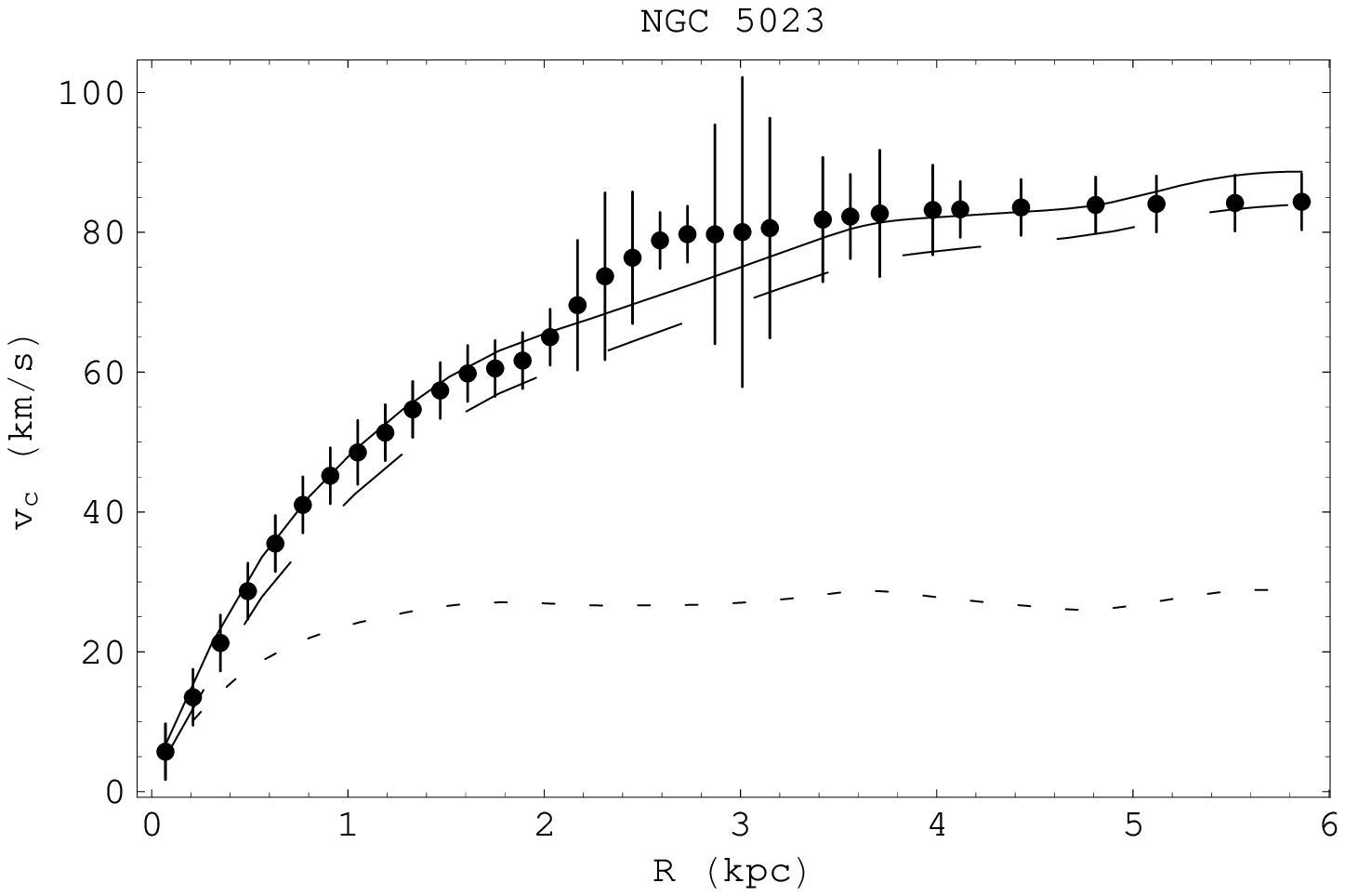}}
\caption{Best fit theoretical rotation curve superimposed to the
data for the LSB galaxy NGC 4455 (left) and NGC 5023 (right). To
better show the effect of the correction to the Newtonian
gravitational potential, we report the total rotation curve
$v_c(R)$ (solid line), the Newtonian one (short dashed) and the
corrected term (long dashed).\label{fig: lsb1}}
\end{figure}
At this point, it is worth wondering whether a link may be found
between $R^n$ gravity and the standard approach based on dark
matter haloes since both theories fit equally well the same data.
 As a matter of fact, it is possible to
define an {\it effective dark matter halo} by imposing that its
rotation curve equals the correction term to the Newtonian curve
induced by $R^n$ gravity. Mathematically, one can split the total
rotation curve derived from $R^n$ gravity as $v_c^2(r) = v_{c,
N}^2(r) + v_{c, corr}^2(r)$ where the second term is the
correction one. Considering, for simplicity a spherical halo
embedding an infinitely thin exponential disk, we may also write
the total rotation curve as $v_c^2(r) = v_{c, disk}^2(r) + v_{c,
DM}^2(r)$ with $v_{c, disk}^2(r)$ the Newtonian disk rotation
curve and $v_{c, DM}^2(r) = G M_{DM}(r)/r$ the dark matter one,
$M_{DM}(r)$ being its mass distribution. Equating the two
expressions, we get\,:

\begin{equation}
M_{DM}(\eta) = 2^{\beta - 5} \eta_c^{-\beta} \pi  (1 - \beta)
\Sigma_0 R_d^2 \eta^{\frac{\beta + 1}{2}} {\cal{I}}_0(\eta, \beta)
\ . \label{eq: mdm}
\end{equation}
with $\eta = r/R_d$, $\Sigma_0 = \Upsilon_{\star} I_0$ and\,:
\begin{equation} {\cal{I}}_0(\eta, \beta) =
\int_{0}^{\infty}{{\cal{F}}_0(\eta, \eta', \beta) k^{3 - \beta}
\eta'^{\frac{\beta - 1}{2}} {\rm e}^{- \eta'} d\eta'} \label{eq:
deficorr}
\end{equation}
with ${\cal{F}}_0$ only depending on the geometry of the system.
Eq.(\ref{eq: mdm}) defines the mass profile of an effective
spherically symmetric dark matter halo whose ordinary rotation
curve provides the part of the corrected disk rotation curve due
to the addition of the curvature corrective term to the
gravitational potential. It is evident that, from an observational
viewpoint, there is no way to discriminate between this dark halo
model and $R^n$ gravity. Having assumed spherical symmetry for the
mass distribution, it is immediate to compute the mass density for
the effective dark halo as $\rho_{DM}(r) = (1/4 \pi r^2)
dM_{DM}/dr$. The most interesting features of the density profile
are its asymptotic behaviors that may be quantified by the
logarithmic slope $\alpha_{DM} = d\ln{\rho_{DM}}/d\ln{r}$ which
can be computed only numerically as function of $\eta$ for fixed
values of $\beta$ (or $n$).  The asymptotic values at the center
and at infinity denoted as $\alpha_0$ and $\alpha_{\infty}$ result
particularly interesting. It turns out that $\alpha_0$ almost
vanishes so that in the innermost regions the density is
approximately constant. Indeed, $\alpha_0 = 0$ is the value
corresponding to models having an inner core such as the cored
isothermal sphere  and the Burkert model \cite{burk}. Moreover, it
is well known that galactic rotation curves are typically best
fitted by cored dark halo models \cite{GS04} therein). On the
other hand, the outer asymptotic slope is between $-3$ and $-2$,
that are values typical of most dark halo models in literature. In
particular, for $\beta = 0.58$ one finds $(\alpha_0,
\alpha_{\infty}) = (-0.002, -2.41)$, which are quite similar to
the value for the Burkert model $(0, -3)$, that has been
empirically proposed to provide a good fit to the LSB and dwarf
galaxies rotation curves. The values of $(\alpha_0,
\alpha_{\infty})$ we find for our best fit effective dark halo
therefore suggest a possible theoretical motivation for the
Burkert\,-\,like models. Now, due to the construction, the
properties of the effective dark matter halo are closely related
to the disk one. As such, we do expect some correlation between
the dark halo and the disk parameters. To this aim, exploiting the
relation between the virial mass and the disk parameters , one can
obtain a relation for the Newtonian virial velocity $V_{vir} = G
M_{vir}/R_{vir}$\,:

\begin{equation}
M_d = \frac{(3/4 \pi \delta_{th} \Omega_m \rho_{crit})^{\frac{1 -
\beta}{4}} R_d^{\frac{1 + \beta}{2}} \eta_c^{\beta}}{2^{\beta - 6}
 (1 - \beta) G^{\frac{5 - \beta}{4}}} \frac{V_{vir}^{\frac{5 -
\beta}{2}}}{{\cal{I}}_0(V_{vir}, \beta)} \label{eq: btfvir} \ .
\end{equation}
We have numerically checked that Eq.(\ref{eq: btfvir}) may be well
approximated as $M_d \propto V_{vir}^{a}$ which has the same
formal structure as the baryonic Tully\,-\,Fisher (BTF) relation
$M_b \propto V_{flat}^a$ with $M_b$ the total (gas + stars)
baryonic mass and $V_{flat}$ the circular velocity on the flat
part of the observed rotation curve. In order to test whether the
BTF can be explained thanks to the effective dark matter halo we
are proposing, we should look for a relation between $V_{vir}$ and
$V_{flat}$. This is not analytically possible since the estimate
of $V_{flat}$ depends on the peculiarities of the observed
rotation curve such as how  far it extends and the uncertainties
on the outermost points. For given values of the disk parameters,
we therefore simulate theoretical rotation curves for some values
of $r_c$ and measure $V_{flat}$ finally choosing the fiducial
value for $r_c$ that gives a value of $V_{flat}$ as similar as
possible to the measured one. Inserting the relation thus found
between $V_{flat}$ and $V_{vir}$ into Eq.(\ref{eq: btfvir}) and
averaging over different simulations, we finally get\,:
$\log{M_b} = (2.88 \pm 0.04) \log{V_{flat}} + (4.14 \pm 0.09)$
while observational data give \cite{ssm}\,:
$\log{M_b} = (2.98 \pm 0.29) \log{V_{flat}} + (3.37 \pm 0.13) \ .$
The slope of the predicted and observed BTF are in good agreement
thus leading further support to our approach. The zeropoint is
markedly different with the predicted one being significantly
larger than the observed one, but it is worth stressing, however,
that both relations fit the data with similar scatter. A
discrepancy in the zeropoint may be due to our approximate
treatment of the effective halo which does not take into account
the gas component. Neglecting this term, we should increase the
effective halo mass and hence $V_{vir}$ which affects the relation
with $V_{flat}$ leading to a higher than observed zeropoint.
Indeed, the larger is $M_g/M_d$, the more the point deviate from
our predicted BTF thus confirming our hypothesis. Given this
caveat, we may therefore conclude with confidence that $R^n$
gravity offers a theoretical foundation even
for the empirically found BTF relation. \\
The results outlined along this paper are referred to a simple
choice of $f(R)$, while it is likely that a more complicated
Lagrangian is needed to reproduce the whole dark sector
phenomenology at all scales. Nevertheless, although not
definitive, these achievements represent an intriguing matter for
future more exhaustive investigations. In particular, exploiting
such models can reveal  a useful approach to  motivate a more
careful search for a single fundamental  theory of gravity able to
explain the full cosmic dynamics with the only two ingredients we
can directly experience, namely the background gravity and the
baryonic matter.
\\ \\
{\bf Acknowledgements} We wish to thank Massimo Capaccioli for the
useful discussions and encouragements which allowed to improve the
paper.


\begin{thebibliography}{99}

\bibitem{capozzcurv}
S. Capozziello, {\it Int. J. Mod. Phys. D}, {\bf 11}, 483 (2002)

\bibitem{cdtt}
S.M. Carroll, V. Duvvuri, M. Trodden, M. Turner, {\it Phys. Rev.
D}, {\bf 70}, 043528, 2004

\bibitem{flanagan}
E.E. Flanagan, {\it Class. Quant. Grav.}, {\bf 21}, 417, 2003

\bibitem{francaviglia}
G. Allemandi, A. Borowiec, M. Francaviglia, {\it Phys. Rev. D},
{\bf 70}, 103503, 2004

\bibitem{odintsovfr}
S. Nojiri, S.D. Odintsov, 2006, hep-th/0601213

\bibitem{birrell}
N.D. Birrell, P.C.W. Davies, {\it Quantum Fields in Curved Space},
Cambridge Univ. Press, Cambridge (UK), 1982


\bibitem{ppnantro}
S. Capozziello, A. Troisi, {\it Phys. Rev. D}, {\bf 72}, 044022,
2005

\bibitem{starobinsky}
A.A. Starobinsky, {\it Phys. Lett. B}, {\bf 91}, 99, 1980

\bibitem{kerner}
R. Kerner, {\it Gen. Rel. Grav.}, {\bf 14}, 453, 1982

\bibitem{noipla}
S. Capozziello, V.F. Cardone,  A. Troisi, 2006, astro-ph/0603522

\bibitem{stelle}
K. Stelle, {\it Gen. Rel. Grav.} {\bf 9}, 353, 1978

\bibitem{mond}
M. Milgrom, {\it Astroph. Journ.}, {\bf 270}, 365, 1983

\bibitem{woodard}
R.P. Woodard, astro-ph/0601672, 2006


\bibitem{curv-ijmpd}
S. Capozziello, V.F. Cardone, S. Carloni, A. Troisi, {\it Int. J.
Mod. Phys. D}, {\bf 12}, 1969, 2003

\bibitem{mimicking}
S. Capozziello, V.F. Cardone, A. Troisi, {\it Phys. Rev. D}, {\bf
71}, 043503, 2005; T. Multamaki and I. Vilja, {\it Phys. Rev. D},
{\bf 73}, 024018, 2006.

\bibitem{dbb02}
W.J.G. de Blok, A. Bosma, {\it Astron. \& Astroph.} {\bf 385},
816, 2002

\bibitem{burk}
A. Burkert,{\it Astroph. Journ.}, {\bf 447}, L25, 1995

\bibitem{GS04}
G. Gentile, P. Salucci, {\it Mon. Not. Roy. Astron. Soc.}, {\it
351}, 953, 2004

\bibitem{ssm}
S.S. McGaugh, {\it Astroph. Journ.}, {\bf 632}, 859, 2005

\end{thebibliography}
\end{document}